\documentclass[onecolumn,amsmath,amssymb,aps,showpacs,superscriptaddress]{revtex4}

\usepackage[dvips]{graphicx}
\usepackage[dvips]{color}
\usepackage{hyperref,breakurl,amsmath,amssymb,pst-node,eepic}
\usepackage{graphicx}% Include figure files
\usepackage{dcolumn}% Align table columns on decimal point
\usepackage{bm}% bold math
\usepackage{hyperref}

\newcommand{\ud}{\mathrm{d}}

\begin{document}

\title{On the non-stationarity of financial time series: impact on optimal portfolio selection}

\author{Giacomo Livan}
\email{glivan@ictp.it}
\affiliation{Abdus Salam International Centre for Theoretical Physics, Strada Costiera 11, 34151 Trieste, Italy}
\author{Jun-ichi Inoue}
\email{j_inoue@complex.ist.hokudai.ac.jp}
\affiliation{Graduate School of Information Science and Technology, Hokkaido University, N14-W9, Kita-ku, Sapporo 060-0814, Japan}
\author{Enrico Scalas}
\email{enrico.scalas@mfn.unipmn.it}
\homepage{www.mfn.unipmn.it/~scalas}
\affiliation{Dipartimento di Scienze e Innovazione Tecnologica, Laboratorio sui Sistemi Complessi, Universit\`a del Piemonte Orientale ``Amedeo Avogadro'', Viale T. Michel 11, 15121 Alessandria, Italy}
\affiliation{BCAM - Basque Center for Applied Mathematics, Bizkaia Technology Park, Building 500, E-48160 Derio, Spain}

\date{\today}

\begin{abstract}
We investigate the possible drawbacks of employing the standard Pearson estimator to measure correlation coefficients between financial stocks in the presence of non-stationary behavior, and we provide empirical evidence against the well-established common knowledge that using longer price time series provides better, more accurate, correlation estimates. Then, we investigate the possible consequences of instabilities in empirical correlation coefficient measurements on optimal portfolio selection. We rely on previously published works which provide a framework allowing to take into account possible risk underestimations due to the non-optimality of the portfolio weights being used in order to distinguish such non-optimality effects from risk underestimations genuinely due to non-stationarities. We interpret such results in terms of instabilities in some spectral properties of portfolio correlation matrices.
\end{abstract}


\maketitle

%%%%%%%%%%%%%%%%%%%%%%%%%%%%%%%%%%%%%%%%%%%%%%%%%%%%%%%%%%%%%%%%%%%%%%%%%%%%%%%%%%%%%%%%%%%%%%%%%%%%%%%%%%
\section{Introduction}
\label{intro}
%%%%%%%%%%%%%%%%%%%%%%%%%%%%%%%%%%%%%%%%%%%%%%%%%%%%%%%%%%%%%%%%%%%%%%%%%%%%%%%%%%%%%%%%%%%%%%%%%%%%%%%%%%

Ever since the fundamental work by Markowitz \cite{Markowitz}, the study and empirical analysis of correlations between stocks traded in a financial market have represented topics of paramount importance in financial analysis. In a nutshell, this is because standard optimal portfolio selection theory (OPST) heavily relies on the knowledge of the portfolio correlation matrix. From the physicist's viewpoint, this poses a measurement problem. Namely, OPST works perfectly whenever the \emph{true} correlations between the stocks forming a portfolio are known. However, such quantities are unobservable by definition, and what one usually does is to measure the correlation coefficient between two stocks $i$ and $j$ according to the well-known Pearson estimator:

\begin{equation} \label{Pearson}
\rho_{ij} = \frac{1}{T} \sum_{t=1}^T r_{it} r_{jt},
\end{equation}
where $r_{it}$ (here assumed to be standardized, \emph{i.e.} zero mean and unit standard deviation) denotes the price change of stock $i$ over the time step from $t-1$ to $t$, for $t = 1,\ldots, T$. It is intuitively clear, and can be rigorously shown, that using longer time series, \emph{i.e.} exploiting more information, will lead to better correlation estimates, so that in the limit of infinitely long time series the Pearson estimator will converge to the true correlation coefficient $\bar{\rho}_{ij}$ between stocks $i$ and $j$: $\rho_{ij} \rightarrow \bar{\rho}_{ij}$ for $T \rightarrow \infty$. Of course, in real financial applications one always needs to cope with time series of finite length. So, when dealing with portfolio selection, being able to make quantitative statements on the reliability of the empirically measured correlation coefficients is highly desirable. In particular, it is relevant to assess the level of measurement noise which might affect correlation estimates and consequently to devise possible filtering techniques to amend empirically observed correlation matrices from noise. Ever since the pioneering works \cite{Laloux2,Plerou}, the Econophyiscs community identified random matrix theory (RMT) as a valuable tool to understand the correlation structure of financial markets (or portfolios) in terms of spectral properties of correlation matrices, shedding light on several interesting stylized facts (see for example \cite{Bouchaud} for a review) and providing rigorous mathematical relations between  a true correlation matrix and its Pearson estimators \cite{Burda2,Burda3}.

The vast majority of the aforementioned results were derived under the assumption that the group of stocks whose correlations are to be measured is described by a joint \emph{stationary} probability distribution. Loosely speaking, this amounts to assuming that a true, \emph{constant}, correlation coefficient $\bar{\rho}_{ij}$ between each pair of stocks $i$ and $j$ actually exists and can be measured, therefore motivating the previously mentioned intuitive notion that employing longer time series provides better correlation estimates. The goal of this paper is to challenge this common knowledge, providing some empirical evidence against it, and to eventually provide a quantitative analysis of the impact that non-stationarities might have on portfolio risk. In recent years, other authors tackled similar problems from different viewpoints. For example, in \cite{Marsili1,Marsili2} a phenomenological model was proposed in order to study the feedback mechanism according to which correlations determine optimal portfolios but are simultaneously affected by investments based on them. Even more recently, different correlation structures induced by non-stationarities in financial time series were exploited to identify and categorize the possible states of a financial market \cite{Guhr}. On a more practical level, it is worth mentioning that the RiskMetrics risk management tool \cite{Mina}, freely available since 1992, introduced a first systematic way to take possible non-stationarities into account by exponentially dampening the contribution of older prices to correlation estimates.

Our approach in this paper will be rather evidence-based, and we shall not try to investigate the possible causes for the emergence of non-stationarities in financial dynamics. Our first goal will be to explore the consequences of possible non-stationarities on empirically measured correlation coefficients between pairs of stocks. More specifically, it is reasonable to expect the distributional properties of a given stock to change over time, especially when considering large time horizons. It is also reasonable to expect different stocks to change at different paces. All in all, the combined effect of rather diverse distributional changes happening at different ``velocities'' will impact the correlations between stocks. In the following we shall look for empirical evidence in this direction, both in a global and local sense. First, we shall consider correlation coefficient measurements (between pairs of stocks) performed over non-overlapping time windows, and we shall perform a statistical test to check their mutual, \emph{global}, compatibility. Then, we shall test the \emph{local} compatibility among consecutive correlation estimates performed over increasingly large time windows. We anticipate here that both tests will reveal substantial violations of the null stationarity hypothesis, and this will motivate our interest in studying the way that such effects might impact the correlation structure of a portfolio and its overall risk.

The paper is organized as follows. The results of the aforementioned global and local stationarity tests, performed on two financial datasets, will be detailed and discussed in Sections \ref{globstat} and \ref{locstat}, respectively. In Section \ref{ops} the effects of inconsistencies in correlation coefficient measurement on portfolio risk assessment will be presented, and a possible interpretation of the results will be given in terms of instabilities in the correlation matrix eigenvalue spectrum. The conclusions and outlook of this work will then be discussed in Section \ref{concl}.

%%%%%%%%%%%%%%%%%%%%%%%%%%%%%%%%%%%%%%%%%%%%%%%%%%%%%%%%%%%%%%%%%%%%%%%%%%%%%%%%%%%%%%%%%%%%%%%%%%%%%%%%%%
\section{Global stationarity}
\label{globstat}
%%%%%%%%%%%%%%%%%%%%%%%%%%%%%%%%%%%%%%%%%%%%%%%%%%%%%%%%%%%%%%%%%%%%%%%%%%%%%%%%%%%%%%%%%%%%%%%%%%%%%%%%%%

\begin{figure}
\begin{center}
\resizebox{0.45\columnwidth}{!}{
  \includegraphics{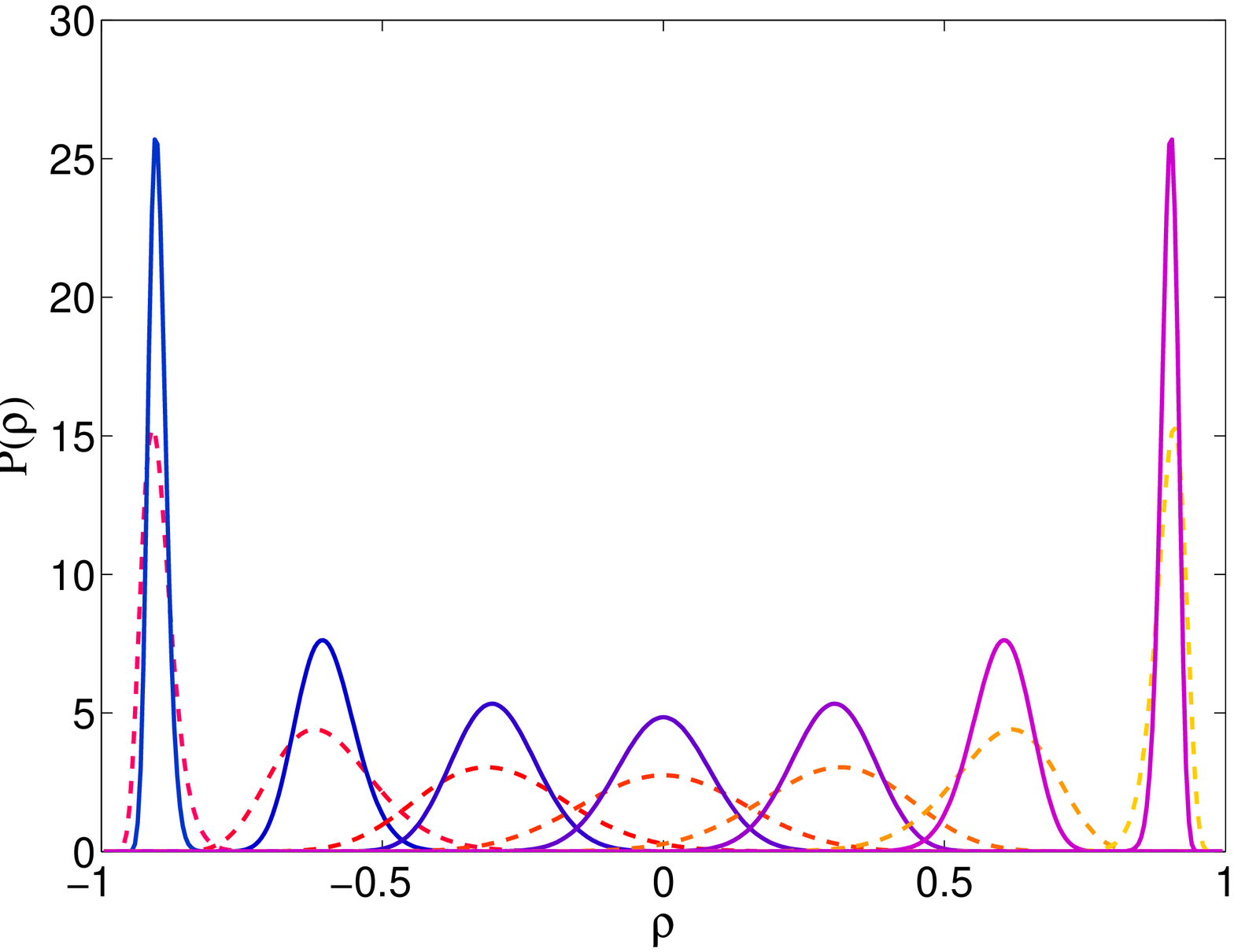}}
\hspace{1cm}
\resizebox{0.45\columnwidth}{!}{
  \includegraphics{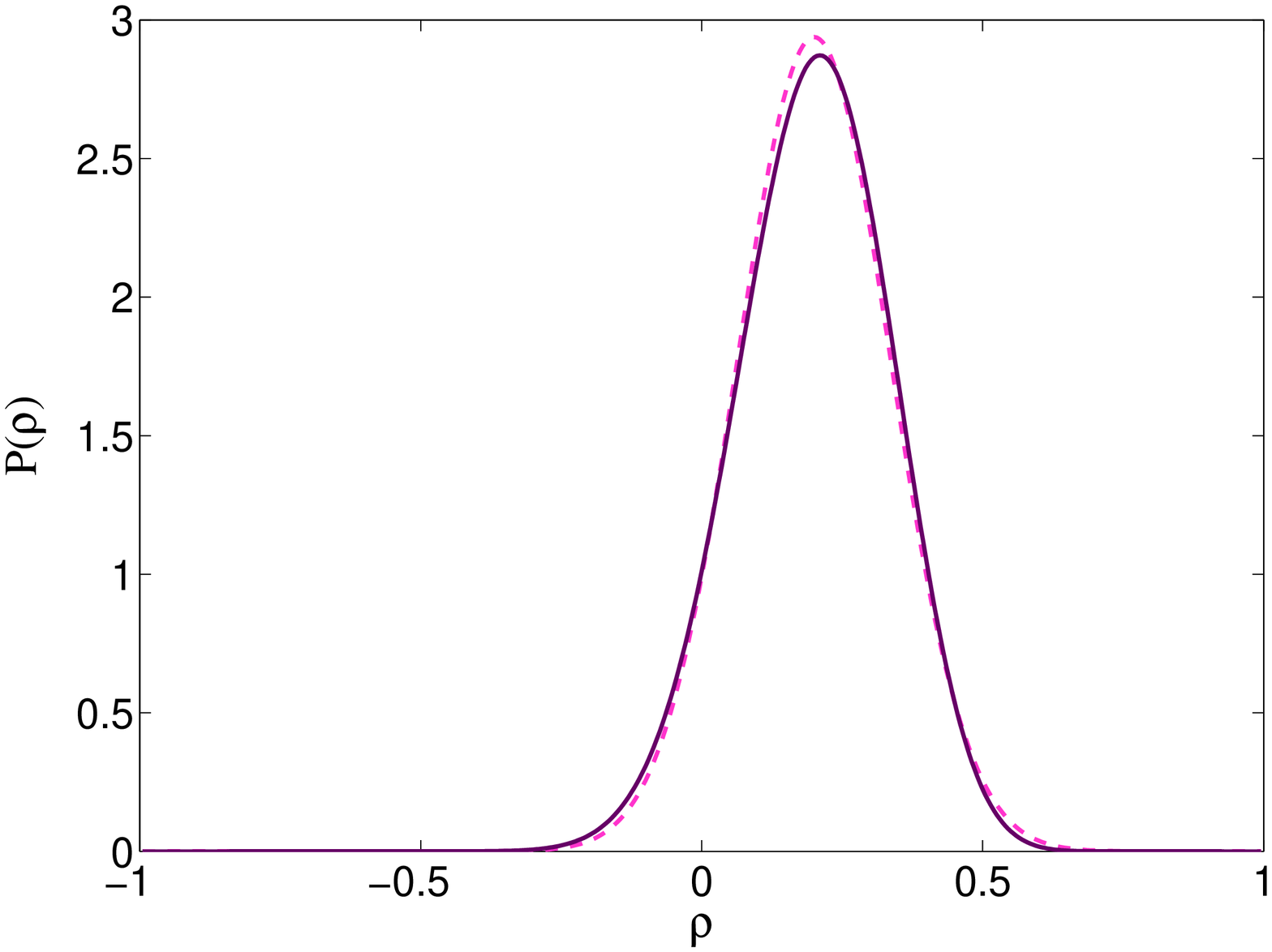}}
\end{center}
\caption{Left: A few examples of the probability density in equation \eqref{rhodistr} for $T=50$ (dashed lines) and $T=150$ (continuous lines). Both sets of curves refer to the values $\bar{\rho} = 0,\pm0.3,\pm0.6,\pm0.9$. As one would naturally expect, it can be seen that, for the same value of $\bar{\rho}$, curves referring to a smaller sample size (\emph{i.e.} a smaller $T$) are broader than their counterparts obtained with a larger $T$. Right: comparison of the probability density in equation \eqref{rhodistr} (with $T = 50$ and $\bar{\rho} = 0.2$) and its Gaussian approximation (dashed line).}
\label{rho_distr_plot}
\end{figure}

The probability distribution of the measured correlation coefficient $\rho$ between a pair of random variables $X$ and $Y$ described by a bivariate Gaussian probability density is given by \cite{Kenney}:

\begin{equation} \label{rhodistr}
P(\rho; \bar{\rho},T) = \frac{1}{\pi} (T-2) (1-\rho^2)^{(T-4)/2} (1-\bar{\rho}^2)^{(T-1)/2} \int_0^{+\infty} \frac{\ud r}{(\mathrm{cosh} \ r - \rho \bar{\rho})^{T-1}} \ , \ \ \ \rho \in [-1,1].
\end{equation}
In this expression $\bar{\rho} \in [-1,1]$ is the true correlation coefficient between $X$ and $Y$, \emph{i.e.} (we denote the expectation with respect to the Gaussian joint probability density describing $X$ and $Y$ as $\mathbb{E}_G[\ldots]$)

\begin{equation} \label{truecorr}
\bar{\rho} = \frac{\mathbb{E}_G[(X-\mathbb{E}_G[X])(Y-\mathbb{E}_G[Y])]}{\sqrt{\mathbb{E}_G[(X-\mathbb{E}_G[X])^2] \mathbb{E}_G[(Y-\mathbb{E}_G[Y])^2]}},
\end{equation}
 whereas $\rho$ is the corresponding Pearson estimator over a sampling time window $T$ (see also equation \eqref{Pearson}), \emph{i.e.}

\begin{equation} \label{Pearson2}
\rho = \frac{1}{T} \sum_{t=1}^T x_t y_t,
\end{equation}
where $x_t$ and $y_t$ ($t = 1, \ldots, T$) are the time-$t$ standardized observations of $X$ and $Y$. In Figure \ref{rho_distr_plot} some examples of the probability density in equation \eqref{rhodistr} are plotted for different values of $\bar{\rho}$ and $T$. It can be shown that the probability density \eqref{rhodistr} has mean and variance given by

\begin{eqnarray} \label{rhomoments}
\mathbb{E}_P[\rho] &=& \bar{\rho} - \frac{\bar{\rho}(1-\bar{\rho}^2)}{2T} \\
\mathbb{E}_P[(\rho-\mathbb{E}_P[\rho])^2] &=& \frac{\left (1-\bar{\rho}^2 \right )^2}{T} \left ( 1 + \frac{11 \bar{\rho}^2}{2T} + \ldots \right ),
\end{eqnarray}
and, already for reasonably large samples, one can realize that such quantities can be safely approximated as $m_P = \bar{\rho}$ and $\sigma^2_P = (1-\bar{\rho}^2)^2/T$, respectively. Moreover, it is
easy to verify that, for sufficiently large sample dimensions $T$, the density \eqref{rhodistr} is fairly well approximated by a Gaussian distribution with mean $m_P$ and standard deviation $\sigma_P$ (see for example the right plot in Figure \ref{rho_distr_plot} for a comparison between the two densities with $T=50$ and $\bar{\rho} = 0.2$), the price to pay being an ``unphysical'' support covering the whole real axis.

The probability density of measured correlation coefficients in equation \eqref{rhodistr} is derived under the assumption that the random variables $X$ and $Y$ are described by a \emph{stationary} joint probability density. In other words, a consequence of this assumption (when verified) is that different Pearson estimates (see equation \eqref{Pearson}) computed over independent, \emph{i.e.} non-overlapping, time windows of the same length $T$ will be distributed according to equation \eqref{rhodistr}. For this very reason, any empirical violation of the probability distribution in equation \eqref{rhodistr} hints at possible non-stationarities in the joint distribution of the random variables $X$ and $Y$, which reflect into non-stationarities in their observed time series $x_t$ and $y_t$. Therefore, a \emph{global} test of stationarity for financial time series can be performed by comparing the empirical distribution of measured correlation coefficients (over non-overlapping time windows) between a given pair of stocks and the probability density in equation \eqref{rhodistr}, seen as a ``null'' stationarity hypothesis, where we assume $\bar{\rho}_{ij}$ to be equal to the correlation coefficient measured over the union of all the non-overlapping time windows. We performed such a test on two financial datasets made of $N_S = 412$ stocks belonging to the American S$\&$P500 Index and $N_F = 137$ stocks belonging to the British FTSE350 Index, respectively. Both datasets are made of 1758 daily price changes covering the years 2005-2011. For each pair of stocks in both datasets ($N_S(N_S-1)/2 = 84666$ pairs for the S$\&$P dataset, $N_F(N_F-1)/2 = 9316$ pairs for the FTSE dataset) the empirical distribution of correlation coefficients estimated over time windows of different lengths was compared to the probability density in equation \eqref{rhodistr} by performing a Kolmogorov-Smirnov (KS) test \cite{Stephens}. The results are summarized in Table \ref{globtest}: as can be seen, especially in the S$\&$P dataset, a relevant fraction of the available stock pairs is found to violate our global 
\begin{table}
\caption{\label{globtest}Results of the KS test comparing the empirical distribution of correlation coefficients measured over non-overlapping time windows of length $T$ to the theoretical one in equation \eqref{rhodistr}. We consider significance values $\alpha = 0.01, 0.05, 0.1$ and time windows of length $T = 25, 50, 100$. For each value of $T$ and for each significance level $\alpha$ the fraction of stock pairs violating the null stationarity hypothesis is reported. alues between parentheses refer to the fractions of stock pairs still violating the test after a random reshuffling is performed in each time series (see the main text for more details). Given that both our S$\&$P and FTSE datasets are made of 1758 returns, when considering samples made of 25 daily returns we obtain 70 correlation coefficient estimates over non-overlapping time windows, when considering $T = 50$ we have 35 estimates and for $T = 100$ we have 17 estimates. Results of the same test (for $\alpha = 0.05$) are also reported for Student-t distributed Monte Carlo data having the same correlation structure and dimensions of the two financial datasets we use (see the main text for more details).}
\begin{ruledtabular}
\begin{tabular}{lccc}
S$\&$P & $T = 25$ & $T = 50$ & $T = 100$ \\
\hline
$\alpha = 0.01$ & 8.49\% (1.87\%) & 4.43\% (0.5\%) & 3.39\% (0.3\%) \\
$\alpha$ = 0.05 & 27.34\% (9.29\%) & 21.04\% (4.94\%) & 18.99\% (1.81\%) \\
$\alpha$ = 0.10 & 42.80\% (25.67\%) & 37.16\% (10.81\%) & 34.45\% (4.57\%) \\
\hline \hline
S$\&$P (Monte Carlo) & $T = 25$ & $T = 50$ & $T = 100$ \\
\hline
$\alpha = 0.05$ & 0.9\% & 1.1\% & 0.8\% \\
\hline \hline
FTSE & $T = 25$ & $T = 50$ & $T = 100$ \\
\hline
$\alpha = 0.01$ & 2.04\% (0.5\%) & 1.28\% (0.3\%) & 0.6\% (0.02\%) \\
$\alpha$ = 0.05 & 10.40\% (3.67\%) & 9.35\% (2.82\%) & 7.01\% (1.5\%) \\
$\alpha$ = 0.10 & 20.03\% (9.35\%) & 19.73\% (4.33\%) & 16.88\% (2.80\%) \\
\hline \hline
FTSE (Monte Carlo) & $T = 25$ & $T = 50$ & $T = 100$ \\
\hline
$\alpha = 0.05$ & 0.8\% & 0.3\% & 0.3\% \\
\end{tabular}
\end{ruledtabular}
\end{table}
stationarity test. In particular, larger violations occur for smaller sample sizes (in our case $T = 25$, roughly corresponding to one trading month): for example, over such sample size we find a remarkably high fraction ($8.49\%$) of stock pairs belonging to the S$\&$P500 which do not fit the stationarity hypothesis even when considering the rather restrictive significance level $\alpha = 0.01$. 

We are by all means aware that performing the tests we have detailed with the probability distribution in equation \eqref{rhodistr} amounts to a certain level of approximation, at least for two reasons. First, equation \eqref{rhodistr} is derived under the assumption of normally distributed data, whereas it is very well-known that daily financial data are usually much more heavy-tailed (typically displaying power law tails with exponent 3 / 3.5). Second, equation \eqref{rhodistr} requires the hypothesis that the random variables under study are described by a multivariate Gaussian distribution, so that a pair of variables has a bivariate Gaussian distribution. This is not exactly the case for financial price changes. Thus, in order to validate the statistical relevance of the results presented in Table \ref{globtest}, we generated two Monte Carlo datasets distributed according to a Student-t distribution having tail exponent equal to 3 and having the same dimensions (both in $N$ and $T$) and correlation structure (over the whole sampling time) of our financial datasets. The results of our global stationarity test, when performed on such synthetic data, are also presented in Table \ref{globtest} for the significance level $\alpha = 0.05$. As the reader can immediately see, we find very few pairs failing the stationarity test: this evidence essentially rules out the heavy-tailed nature of the data and their distributional properties as possible causes for the very large fraction of stock pairs failing the test. It is then safe to state that non-stationarities are indeed the main failure cause in such a test. 

As a further check of the statistical significance of the results presented in Table \ref{globtest}, we also performed the same global stationarity test on our two datasets after performing a synchronous random reshuffling. More precisely, in each of the two datasets we performed \emph{one same} random reshuffling of the price changes in each time series. Such an operation clearly leaves the overall cross-correlation structure of the dataset intact, whereas it destroys (most of) the correlation dynamics in the data. In other words, this type of reshuffling is supposed to generate a stationary dataset. The results we obtained from our global stationarity test after performing the reshuffling are reported in Table \ref{globtest} within parentheses. As one can see, for each combination of the sampling time $T$ and the significance level $\alpha$, the fraction of stock pairs failing the test is significantly reduced when the reshuffling is performed.

In the following Section we shall try to motivate these findings by investigating more ``local'' properties of empirically measured correlation coefficients.

%%%%%%%%%%%%%%%%%%%%%%%%%%%%%%%%%%%%%%%%%%%%%%%%%%%%%%%%%%%%%%%%%%%%%%%%%%%%%%%%%%%%%%%%%%%%%%%%%%%%%%%%%%
\section{Local stationarity}
\label{locstat}
%%%%%%%%%%%%%%%%%%%%%%%%%%%%%%%%%%%%%%%%%%%%%%%%%%%%%%%%%%%%%%%%%%%%%%%%%%%%%%%%%%%%%%%%%%%%%%%%%%%%%%%%%%

\begin{figure}
\begin{center}
\resizebox{0.45\columnwidth}{!}{
  \includegraphics{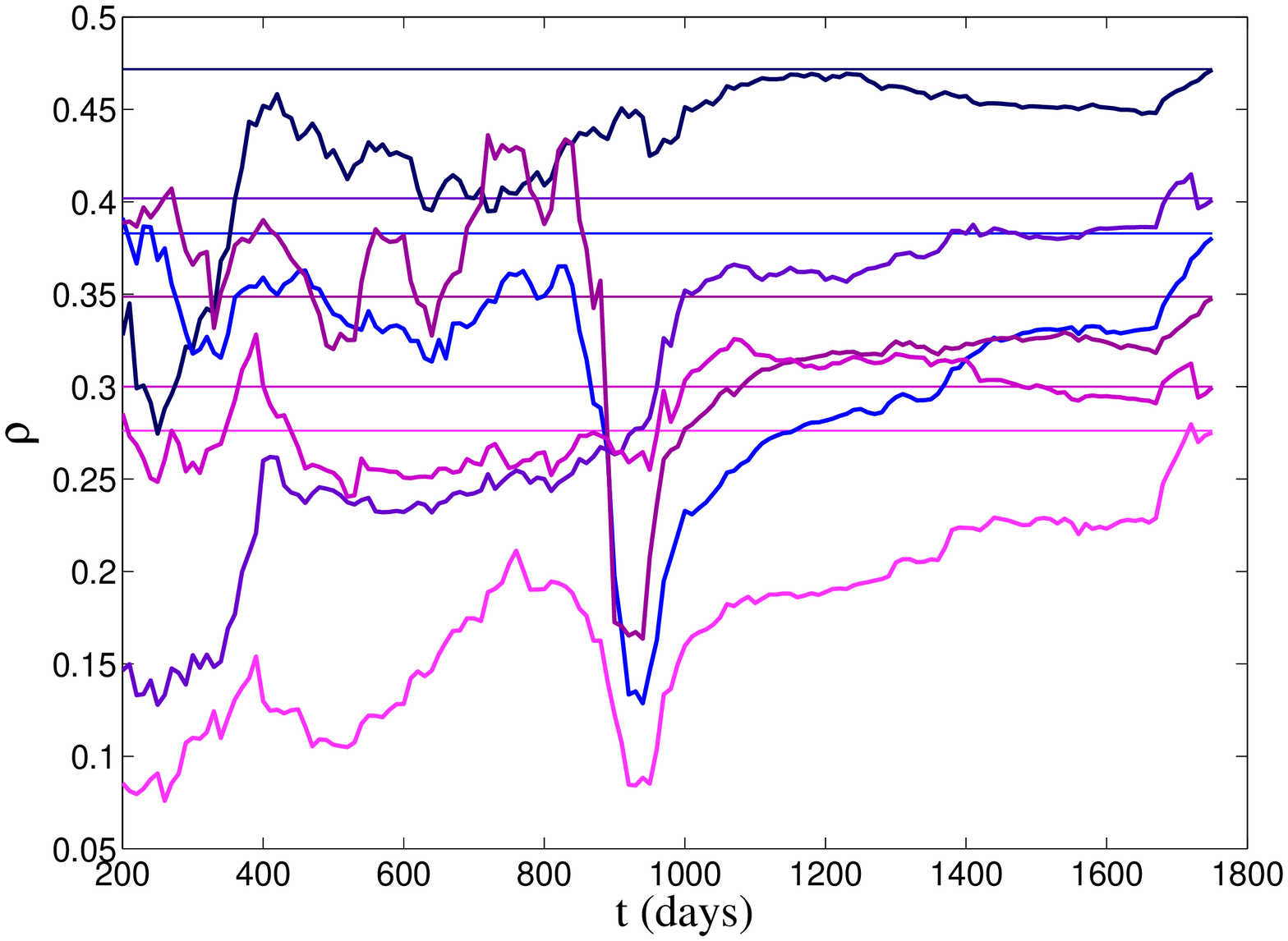}}
\hspace{1cm}
\resizebox{0.45\columnwidth}{!}{
  \includegraphics{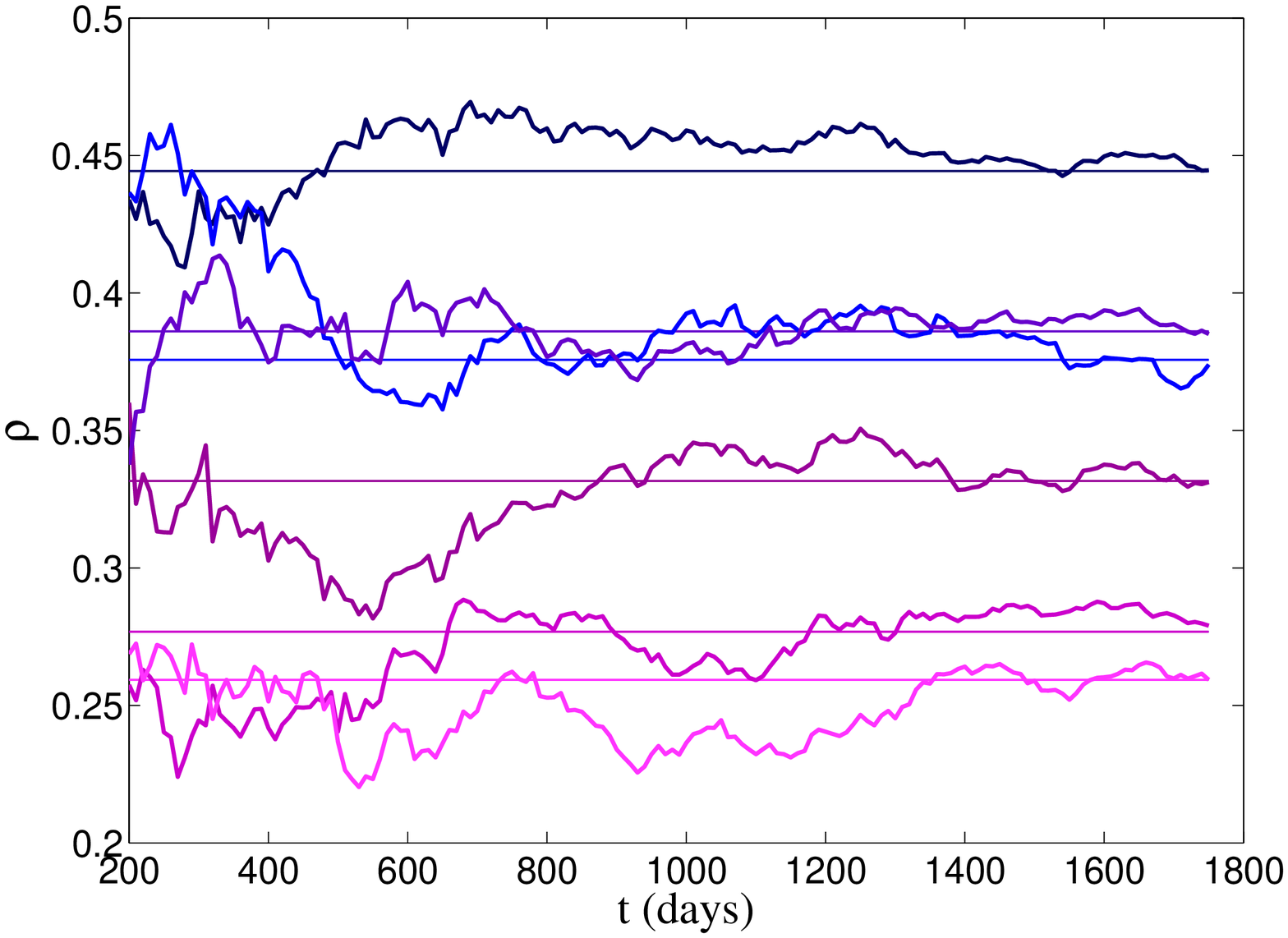}}
\end{center}
\caption{Both plots show the ``time evolution'' of correlation estimates computed over increasingly large time windows. The left plot refers to the correlation coefficients computed for a group of four stocks belonging to our S$\&$P dataset, while the right plot refers to Monte Carlo data generated with the same correlation structure of the stocks in the left plot. Correlation coefficient estimates computed over the whole available time window (1758 days) are shown as horizontal lines.}
\label{corrconv}
\end{figure}

In Figure \ref{corrconv} a possible qualitative explanation for the evidence presented in the previous Section (see Table \ref{globtest} in particular) is pictured by plotting the ``time evolution'' of the correlation coefficient estimates between
four stocks (six correlation coefficients) belonging to the S$\&$P dataset. More specifically, starting from a first estimate computed over the first 200 days of the 1758 making the dataset, correlation coefficients among price changes are recomputed over increasingly large time windows by adding 10 days each time. As can be seen in the left plot of Figure \ref{corrconv}, the correlation estimates display a rather wild fluctuating behavior, and, for most of the time, have actually very little to do with the estimates computed over the whole period of 1758 days (shown as straight lines in the plot). If a constant, or at least approximately constant, true  value of the correlation coefficient between two stocks actually existed, then the estimation process would move on as in the right plot of Figure \ref{corrconv}, which was produced with simulated data having the same correlation structure of the four stocks in the left plot. In this case, a true correlation coefficient does exist, and, as a matter of fact, estimates obtained from larger samples tend to lie closer to such a true value, progressively reducing the fluctuations around it (an infinitely long time series would eventually lead the correlation coefficient estimates to converge exactly to their corresponding true value). In this respect, the two plots in Figure \ref{corrconv} drastically defy the common knowledge that longer time series should eventually produce better correlation estimates.

It is easy to verify that the statistical error on the correlation coefficient estimates introduced in equation \eqref{Pearson} scales as $T^{-1/2}$. Thus, a \emph{local} stationarity test can be performed as follows. Let us divide the whole sampling interval $T$ into $K$ equal parts of size $\tau$, \emph{i.e.} $T = K\tau$. So, given the price changes $r_{it}$ and $r_{jt}$ of two stocks, let us estimate the correlation coefficient between them up to the $\kappa$-th interval ($\kappa = 1, \ldots, K-1$) by specializing equation \eqref{Pearson} as follows:

\begin{equation} \label{kPearson}
\rho_\kappa = \frac{1}{\kappa \tau} \sum_{t=1}^{\kappa \tau} r_{it} r_{jt}.
\end{equation}
Then, one can compute the estimate $\rho_{\kappa+1}$ over the next interval. Since the error on $\rho_\kappa$ is of order $\sigma_\kappa = 1/\sqrt{\kappa \tau}$, one can then assess the compatibility of consecutive correlation estimates between the stocks $i$ and $j$ by checking whether $\rho_{\kappa+1} \in [\rho_\kappa - n \sigma_\kappa, \rho_\kappa + n \sigma_\kappa]$ for some integer $n$ or not.

The aforementioned local stationarity test was performed on the previously introduced S$\&$P and FTSE datasets. In particular, for each pair of stocks $i$ and $j$ we computed a first correlation estimate over $T_1 < T$ days, and then divided the remaining $T_2 = T - T_1$ days into $K$ chunks of length $\tau$, as in the previous example. The following cases were considered: $T_1 = 200$ and $\tau = 50$ (giving 32 correlation estimates for each pair of stocks), $T_1 = 200$ and $\tau = 100$ (16 estimates), $T_1 = 250$ and $\tau = 250$ (7 estimates). The results are summarized in Table \ref{loccorr}: the test was performed, for all of the
\begin{table}
\caption{\label{loccorr} Results of the local stationarity test described in the main test. For each dataset (S$\&$P and FTSE) we report the number of correlation coefficient estimates failing the test for different confidence interval amplitudes characterized by $n = 1,3,5$. We also report the same results for two Monte Carlo datasets generated with the same correlation structure of the original financial datasets.}
\begin{ruledtabular}
\begin{tabular}{lccc}
S$\&$P & $\tau = 50$ & $\tau = 100$ & $\tau = 250$ \\
\hline
$n = 1$ & 8.69\% & 19.32\% & 40.39\% \\
$n = 3$ & 1.00\% & 2.75\% & 8.85\% \\
$n = 5$ & 0.1\% & 0.3\% & 1.20\% \\
\hline \hline
S$\&$P (Monte Carlo) & $\tau = 50$ & $\tau = 100$ & $\tau = 250$ \\
\hline
$n = 1$ & 1.1\% & 2.74\% & 7.43\% \\
$n = 3$ & 0\% & 0.1\% & 0.2\% \\
$n = 5$ & 0\% & 0\% & 0\% \\
\hline \hline
FTSE & $\tau = 50$ & $\tau = 100$ & $\tau = 250$ \\
\hline
$n = 1$ & 9.52\% & 21.46\% & 39.65\% \\
$n = 3$ & 0.7\% & 2.37\% & 5.74\% \\
$n = 5$ & 0.2\% & 0.5\% & 1.04\% \\
\hline \hline
FTSE (Monte Carlo) & $\tau = 50$ & $\tau = 100$ & $\tau = 250$ \\
\hline
$n = 1$ & 1.06\% & 8.96\% & 17.03\% \\
$n = 3$ & 0\% & 1.02\% & 0.7\% \\
$n = 5$ & 0\% & 0\% & 0\% \\
\end{tabular}
\end{ruledtabular}
\end{table}
aforementioned values of the lag $\tau$, by counting the fraction of all correlation estimates computed over the $(\kappa+1)$-th lag falling outside the interval $[\rho_\kappa - n \sigma_\kappa, \rho_\kappa + n \sigma_\kappa]$, for $n = 1,3,5$. As one can see, we find relevant fractions of estimates violating our local stationarity test, especially when considering $\tau = 250$, roughly corresponding to one trading year. In this latter case we find a remarkable $1.20\%$ of the S$\&$P estimates and $1.04\%$ of the FTSE estimates failing the test even when $n = 5$, meaning that year-to-year correlation estimates between some stock pairs can exhibit unexpectedly huge fluctuations. In analogy to what we did in the previous Section (see Table \ref{globtest}), we verified the statistical relevance of our findings by performing the local stationarity test on two sets of synthetic Student-t distributed data generated with the same correlation structure (over the whole time window $T = T_1 + T_2$) of our original S$\&$P and FTSE datasets. As can be seen in Table \ref{loccorr}, the fraction of correlation estimates failing the test is significantly reduced when considering Monte Carlo data. This result essentially rules out the data distributional properties, such as heavy tails, as possible causes for failing the local stationarity test, leaving us with non-stationarities as the main possible cause.

%%%%%%%%%%%%%%%%%%%%%%%%%%%%%%%%%%%%%%%%%%%%%%%%%%%%%%%%%%%%%%%%%%%%%%%%%%%%%%%%%%%%%%%%%%%%%%%%%%%%%%%%%%
\section{Optimal portfolio selection}
\label{ops}
%%%%%%%%%%%%%%%%%%%%%%%%%%%%%%%%%%%%%%%%%%%%%%%%%%%%%%%%%%%%%%%%%%%%%%%%%%%%%%%%%%%%%%%%%%%%%%%%%%%%%%%%%%

According to OPST, due to Markowitz \cite{Markowitz}, the optimal weights for a portfolio built out of a given set of stocks can be determined explicitly in a number of different situations. Let us assume that our portfolio has to be put together as a combination of $N$ stocks, and let us indicate, as already done throughout the paper, the time-$t$ price changes of such stocks as $r_{it}$. Thus, assuming we hold a quantity $w_i$ of stock $i$ over the time interval $[t-1,t]$, the portfolio value change over such interval is simply given by

\begin{equation} \label{portfoliopc}
\Pi_t = \sum_{i=1}^N w_i r_{it}.
\end{equation}
Usually, Markowitz's OPST is formulated in terms of portfolio return maximization for a given fixed risk level, quantified by the portfolio variance, or risk minimization for a fixed expected return. The portfolio variance is given by

\begin{equation} \label{portfoliorisk}
\sigma^2_\Pi = \sum_{i,j=1}^N w_i w_j C_{ij} = \sum_{i,j=1}^N w_i w_j \sigma_i \sigma_j \rho_{ij},
\end{equation}
where $C_{ij} = \sigma_i \sigma_j \rho_{ij}$ is the covariance matrix element between stocks $i$ and $j$, whereas $\sigma_i$ is the standard deviation of $r_i$ and $\rho_{ij}$ is, as usual, the correlation coefficient between stocks $i$ and $j$. Following \cite{PK}, one can simply choose to minimize portfolio risk under a budget constraint:

\begin{equation} \label{constr}
\sum_{i=1}^N w_i = 1.
\end{equation}
So, all in all, this amounts to solving the following optimization problem:

\begin{equation} \label{optprob}
\frac{\partial }{\partial w_i} \left [ \sigma^2_\Pi + \xi \left ( \sum_{j=1}^N w_j - 1 \right ) \right ]_{w_i = w_i^*} = 0,
\end{equation}
where $w_i^*$ denotes the optimal weight on stock $i$ and $\xi$ is a Lagrange multiplier. Quite straightforwardly, one finds

\begin{equation} \label{optweight}
w_i^* = \frac{\sum_{j=1}^N (C^{-1})_{ij}}{\sum_{j,k=1}^N (C^{-1})_{jk}},
\end{equation}
and this result shows that optimal portfolio weights heavily depend on correlations through the covariance matrix $C_{ij} = \sigma_i \sigma_j \rho_{ij}$. Incidentally, let us remark here that the above construction actually implies $T > N$, where $T$ is the number of recorded price changes for each stock. In a nutshell, this is because the eigenvalue spectrum of the covariance matrix $\mathbf{C}$ yields $\min \{ N,T \}$ non-zero eigenvalues. Now, since $\mathbf{C}$ is $N \times N$, this means that if one had $T < N$, then $N-T$ zero modes would appear in the spectrum, and, given the dependence of the $w_i^*$s on the inverse of the covariance matrix, this would make the optimal portfolio problem ill-defined. \\

As discussed in \cite{PK,PKN,Laloux}, different notions of risk can be introduced from the optimal weights in \eqref{optweight}, depending on the covariance matrix being used. First of all, assuming correlations and volatilities to be constant over time, one can define the \emph{true} optimal risk of the portfolio $\Pi$:

\begin{equation} \label{truerisk}
(\sigma_\Pi^{\mathrm{T}})^2 = \sum_{i,j=1}^N C_{ij}^{\mathrm{T}} \ (w_i^\mathrm{T})^* (w_j^\mathrm{T})^*,
\end{equation}
where $\mathbf{C}^{\mathrm{T}}$ (with entries $C_{ij}^{\mathrm{T}}$) is the true portfolio covariance matrix, whereas $(w_i^\mathrm{T})^*$ denotes the corresponding optimal weights. Of course, the true risk of a portfolio cannot be known, and the only thing one can do is to compute the optimal weights starting from a covariance matrix estimate $\mathbf{C}^\mathrm{E}$, computed over some interval $T$. So, the estimated risk from a given sample reads

\begin{equation} \label{insample}
(\sigma_\Pi^{\mathrm{E}})^2 = \sum_{i,j=1}^N C_{ij}^{\mathrm{E}} \ (w_i^\mathrm{E})^* (w_j^\mathrm{E})^*.
\end{equation}
This type of risk can also be meaningfully called \emph{in-sample} risk, since it only provides risk estimates over past time intervals. What can actually be done in practice in order to infer future risk levels is to estimate the covariance matrix over some time interval $T_1$, compute the corresponding optimal weights via equation \eqref{optweight}, and then retain such weights over a consecutive time interval $T_2$. This is called \emph{realized} risk, and it reads

\begin{equation} \label{outofsample}
(\sigma_\Pi^{\mathrm{R}})^2 = \sum_{i,j=1}^N C_{T_2,ij}^{\mathrm{E}} \ (w_{T_1,i}^\mathrm{E})^* (w_{T_1,j}^\mathrm{E})^*,
\end{equation}
where $w_{T_1,i}^\mathrm{E}$ is the optimal weight computed over time interval $T_1$, while $\mathbf{C}_{T_2}^\mathrm{E}$ (with entries $C_{T_2,ij}^\mathrm{E}$) is the covariance matrix over the following time interval $T_2$. So, all in all, the main idea is to use the in-sample risk computed over $T_1$ as a proxy for the realized risk over $T_2$. In references \cite{PK} the ratio 

\begin{equation} \label{qratio}
q = \sigma_\Pi^\mathrm{R} / \sigma_\Pi^\mathrm{E},
\end{equation}
which gives information on possible underestimations of the realized risk, was extensively studied by means of Monte Carlo simulations (so that also the true portfolio risk \eqref{truerisk} was known and under control). A fundamental and intuitive result presented in \cite{PK} is that the $q$ ratio \eqref{qratio} is always bigger than one. This is due to the fact that $\sigma_\Pi^\mathrm{E}$ is a genuinely optimal risk, whereas $\sigma_\Pi^\mathrm{R}$ is not, being computed with the weights $w_{T_1,i}^\mathrm{E}$, which are not optimal with respect to the covariance matrix $\mathbf{C}_{T_2}^\mathrm{E}$. Another, much less intuitive, result presented in \cite{PK} is that the $q$ ratio is essentially model-independent. In other words, whenever the number of stocks $N$ and the sampling intervals $T_1$ and $T_2$ are kept fixed, the $q$ ratio and its sample-to-sample fluctuations show no real dependence on the type of model used to define the true covariance matrix $\mathbf{C}^\mathrm{T}$, which obviously determines the nature of its estimators $\mathbf{C}_1^\mathrm{E}$ and $\mathbf{C}_2^\mathrm{E}$. See also \cite{Galluccio,Burda,PPNK} for similar discussions on related topics.

The framework outlined above allows to quantify, via the $q$ ratio in equation \eqref{qratio}, the risk underestimation being introduced whenever past optimal weights are retained over future time intervals. Moreover, as already pointed out, even the statistical error on $q$ is model-independent up to very good approximation. So, whenever $N$, $T_1$ and $T_2$ have been fixed, one can obtain a very accurate confidence level interval for $q$ by performing a portfolio Monte Carlo simulation with statistics large enough to reach the desired accuracy. We relied on these results in order to check whether secondary effects due to
\begin{figure}
\begin{center}
\resizebox{0.45\columnwidth}{!}{
  \includegraphics{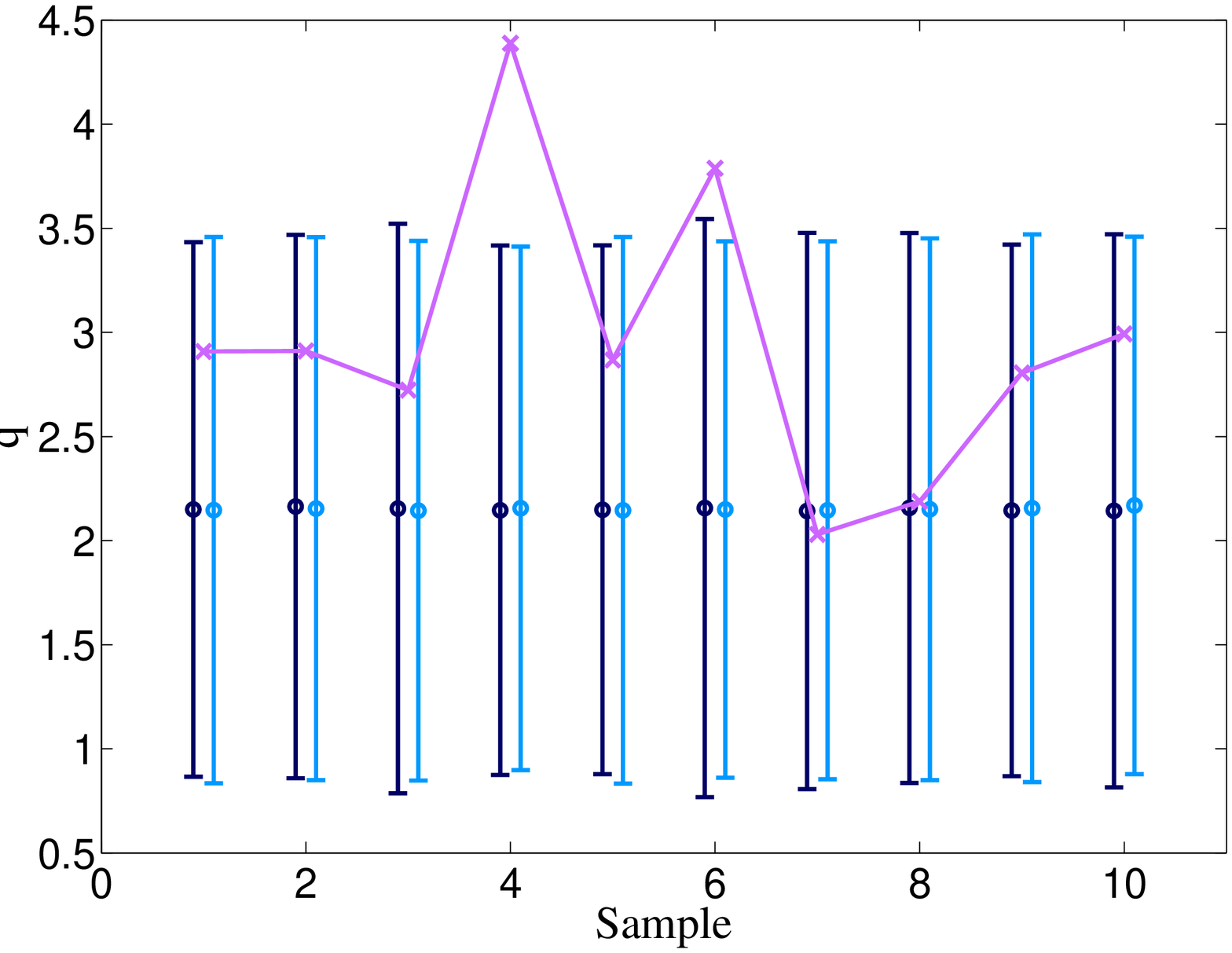}}
\hspace{1cm}
\resizebox{0.45\columnwidth}{!}{
  \includegraphics{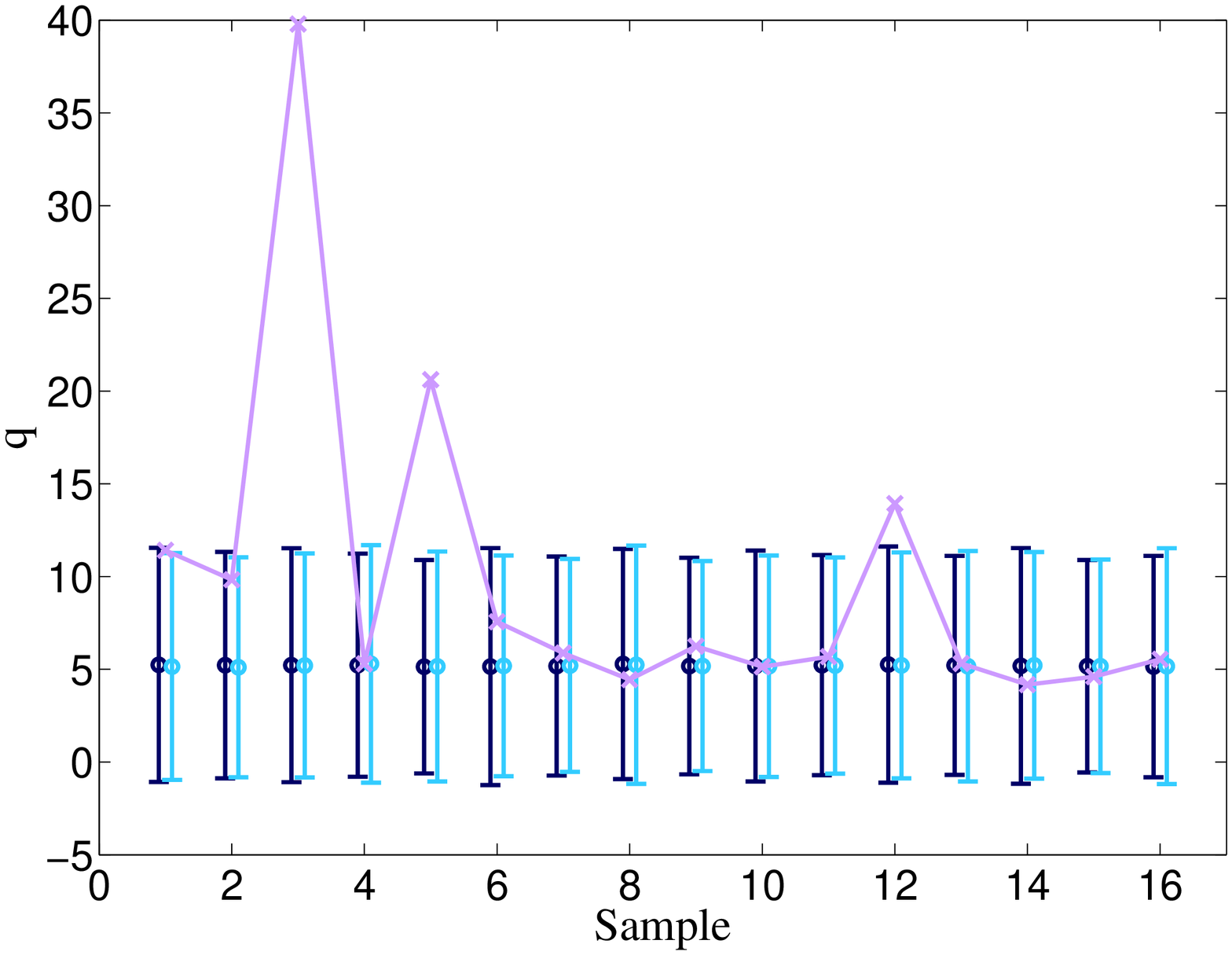}}
\end{center}
\caption{Comparison between the $q$ ratios (equation \eqref{qratio}) obtained with real portfolio data (purple crosses) and the corresponding 5-sigma confidence level intervals obtained under the hypothesis of stationary data (light blue for uncorrelated Gaussian data and dark blue for Gaussian data with the same correlation structure of the empirical data in use).}
\label{portfolios}
\end{figure}
non-stationarities might lead to further risk underestimations, \emph{i.e.} to unexpectedly large values of the $q$ ratio in \eqref{qratio}. In Figure \ref{portfolios} we report examples of the results we obtained. We randomly selected $N = 80$ stocks both in the S$\&$P and FTSE datasets, and used them to form portfolios (we checked that random stock selection does not impact the overall qualitative appearance of the results). Then, in the S$\&$P case we divided the time series for the $N$ selected stocks into chunks of length $T_1 = T_2 = 150$ days, using the first $T_1$ days for determining optimal portfolio weights, and then retaining them over the second period of $T_2$ days. We did the same thing for the FTSE dataset using $T_1 = T_2 = 100$ days in that case. In order to work with as many samples as possible, we employed the $T_2$ days used to compute the realized risk of the $n$-th portfolio as the $T_1$ days used to compute the in-sample risk of the $(n+1)$-th portfolio. The dots and error bars in Figure \ref{portfolios} refer to Monte Carlo simulations of 100 synthetic portfolios generated with the same covariance structure (dark blue) of the data they are compared to and with no correlation at all (light blue). The dots represent estimates for the $q$ ratio, whereas bars represent 5-standard deviation intervals (incidentally, one can see there is almost no difference between the two intervals, due to the aforementioned model-independence of the $q$ ratio and its error). On the other hand, purple crosses represent $q$ ratios computed with real data. The plot on the left is obtained from S$\&$P data, whereas the plot on the right refers to FTSE data. As can be seen, especially in this latter case, surprisingly large risk underestimations can happen, well outside the range one would have simply because of the non-optimality of the portfolio weights being held over $T_2$. From now on we shall refer to such region as to the non-optimality region. In the following we shall see how this evidence can be interpreted by investigating the spectral properties of portfolio correlation matrices. However, before we do so, let us clarify a subtle (yet rather important) point. Back in Section \ref{globstat}, it was shown that correlation estimates computed over non-overlapping time windows are not always compatible with each other, and this result was essentially interpreted, in Section \ref{locstat}, in terms of time series non-stationarities and the consequently induced correlation dynamics (portrayed by the left plot in Figure \ref{corrconv}). We commented these observations by remarking that such evidence goes against the intuitive notion that measuring correlations over very long time series produces better estimates. However, such a remark is not to be confused with another effect related to the use of time series of different length when assessing portfolio risk. Such an effect is portrayed in Figure \ref{qplots}, where the $q$ ratio and its error are evaluated for the same S$\&$P portfolio used to produce the left plot in Figure \ref{portfolios} with $T_2 = 150$ and $T_1 = 100,125,150$ (red, blue and purple lines, respectively). As one can immediately recognize, in this case employing longer time series generally reduces the $q$ ratio and its statistical uncertainty. This means that employing longer time series reduces the possible risk underestimations due to the non-optimality of the portfolio weights being used. However, it is important to stress that this does not necessarily protect against unexpectedly large fluctuations due to non-stationarities, as exemplified by sample 4 in Figure \ref{qplots}, which violates the non-optimality bounds for all three values of $T_1$.
\begin{figure}
\begin{center}
\resizebox{0.45\columnwidth}{!}{
  \includegraphics{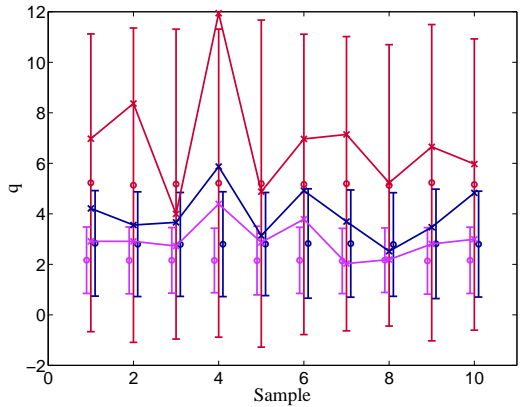}}
\end{center} 
\caption{$q$ ratio for the same portfolio used to produce the left plot in Figure \ref{portfolios}. For all samples we have $T_2 = 150$ days. The lengths of the sampling interval $T_1$ are 100 (red lines), 150 (blue lines) and 200 (purple line) days.}
\label{qplots}
\end{figure}

%%%%%%%%%%%%%%%%%%%%%%%%%%%%%%%%%%%%%%%%%%%%%%%%%%%%%
\subsection{Spectral analysis}
%%%%%%%%%%%%%%%%%%%%%%%%%%%%%%%%%%%%%%%%%%%%%%%%%%%%%

Given a set of $N$ stocks and two consecutive sampling periods $T_1$ and $T_2$, let us introduce the $N \times N$ correlation matrices $\boldsymbol{\rho}_{T_1}^\mathrm{E}$ and $\boldsymbol{\rho}_{T_2}^\mathrm{E}$ with entries $\rho_{T_1,ij}^\mathrm{E}$ and $\rho_{T_2,ij}^\mathrm{E}$ representing the estimated Pearson correlation coefficient between the price changes of stocks $i$ and $j$ over the sampling periods $T_1$ and $T_2$, respectively. Correlation matrices are symmetric and positive definite by construction. Thus, we can introduce the eigenvalues $0 \leq \lambda_{T_m,1} \leq \lambda_{T_m,2} \leq \ldots \leq \lambda_{T_m,N}$ (where $m = 1,2$ denotes the sampling period) of the two matrices $\boldsymbol{\rho}_{T_1}^\mathrm{E}$ and $\boldsymbol{\rho}_{T_2}^\mathrm{E}$ introduced above, and the corresponding normalized eigenvectors $\mathbf{v}_{T_m,n} = (v_{T_m,n}^{(1)}, \ldots, v_{T_m,n}^{(N)})$ (where $m = 1,2$ and $n = 1, \ldots, N$). Now, ever since the works \cite{Laloux2,Plerou} it has been very well known that the eigenvalue spectrum of a financial correlation matrix typically displays a main bulk of small eigenvalues plus a few, much larger, eigenvalues ``leaking out'' of such bulk. As a first approximation, the former can be identified as the noisy part of the spectrum, whereas the latter can be seen as the only eigenvalues carrying meaningful information on the correlation structure of the market or portfolio under analysis (see \cite{Livan}, where a more refined analysis of the information amount actually carried by the bulk and the larger eigenvalues was carried out). Moreover, taking a closer look at the large eigenvalues  (for example by means of principal component analysis (PCA)) unravels the following properties. The largest eigenvalue $\lambda_{T_m,N}$ is usually of order $N$, and it emerges as a consequence of global fluctuations involving all the stocks in the market or portfolio. Thus, it is generally labeled as the \emph{market eigenvalue}. On the other hand, the next few largest eigenvalues $\lambda_{T_m,N-1}, \lambda_{T_m,N-2}, \ldots \lambda_{T_m,N-S}$ (usually for $S \sim 3-5$) typically arise from fluctuations involving only a certain fraction of the $N$ stocks under consideration, and such stocks are found to belong to the same market sector in most cases. For this very reason, the eigenvalues $\lambda_{T_m,N-1}, \lambda_{T_m,N-2}, \ldots, \lambda_{T_m,N-S}$ are generally labeled as \emph{sector eigenvalues}.

Such evidence can be often unraveled by means of inverse participation ratios (IPRs). Given an $N-$dimensional normalized vector $\mathbf{w} = (w^{(1)},\ldots,w^{(N)})$, its IPR $I_\mathbf{w}$ is defined as the sum of the fourth power of its components:

\begin{equation} \label{ipr}
I_\mathbf{w} = \sum_{i=1}^N \left ( w^{(i)} \right )^4.
\end{equation}
So, IPRs can essentially be regarded as a tool to highlight the localization properties of a normalized vector. As a matter of fact, a fully delocalized vector with all components equal to $1/\sqrt{N}$ has IPR equal to $1/N$, while a completely localized vector with one component equal to one and all others equal to zero has IPR equal to one. Moreover, PCA tells us that the generic price change $r_{T_m,i}(t)$ (for $m = 1,2$, $i=1,\ldots,N$ and time $t \in (1,T_1]$ for $m=1$ or $t \in [T+1,T]$ for $m=2$) can be decomposed as

\begin{equation} \label{PCA}
r_{T_m,i}(t) = \sum_{\ell=1}^N \sqrt{\lambda_{T_m,\ell}} \ v_{T_m,\ell}^{(i)} \ e_{T_m,\ell}(t),
\end{equation}
where $e_{T_m,\ell}$ denotes the $\ell$-th principal component. Equation \eqref{PCA} shows that each eigenvector enters the dynamics of each stock. Therefore, a very delocalized eigenvector, as the one typically related to the market eigenvalue, will make sure that the corresponding principal component drives all stocks approximately in the same way. Conversely, a highly localized eigenvector will cause the corresponding principal component to impact only very few variables (as a matter of fact eigenvectors related to sector eigenvalues are usually rather localized). In the light of these considerations, let us now see how the spectral analysis of the correlation matrices $\boldsymbol{\rho}_{T_1}^\mathrm{E}$ and $\boldsymbol{\rho}_{T_2}^\mathrm{E}$ can help to understand what causes the large spikes of the $q$ ratio shown in Figure \ref{portfolios}.

In Tables \ref{SP_portfolio} and \ref{FTSE_portfolio} a few spectral quantities of interest related to the plots in
\begin{table}
\caption{\label{SP_portfolio} Spectral quantities of interest (market eigenvalues, sum of the first few sector eigenvalues and IPR of the market eigenvector) computed from the correlation matrix of a portfolio made of $N = 80$ stocks belonging to the S$\&$P dataset. The sampling times $T_1$ and $T_2$ are both equal to 150 days.}
\begin{ruledtabular}
\begin{tabular}{lcccccc}
Sample & $\lambda_{T_1,N}$ & $\lambda_{T_2,N}$ & $\Lambda_{T_1}$ & $\Lambda_{T_2}$ & $I_{\mathbf{v}_{T_1,N}} (\times 10^{-2})$ & $I_{\mathbf{v}_{T_2,N}} (\times 10^{-2})$ \\
\hline
    1 & 19.91 & 17.20 & 11.29 & 11.90 & 1.51 & 1.55 \\
    2 & 17.20 & 20.10 & 11.90 & 12.43 & 1.55 & 1.48 \\
    3 & 20.10 & 19.47 & 12.43 & 10.42 & 1.48 & 1.47 \\
    $4^*$ & 19.47 & 30.15 & 10.42 &  9.46  & 1.47 & 1.44 \\
    5 & 30.15 & 27.81 &  9.46  &  12.88 & 1.44 & 1.55 \\
    $6^*$ & 27.81 & 43.87 &  12.88 &  9.40  & 1.55 & 1.32 \\
    7 & 43.87 & 36.11 &  9.40   &  9.87  & 1.32 & 1.44 \\
    8 & 36.11 & 31.91 &  9.87   &  7.86  & 1.44 & 1.42 \\
    9 & 31.91 & 45.30 &  7.86   &  5.76  & 1.42 & 1.31 \\
  10 & 45.30 & 28.44 &  5.76   &  9.61 &  1.31 & 1.42 \\
\end{tabular}
\end{ruledtabular}
\end{table}
\begin{table}
\caption{\label{FTSE_portfolio} Spectral quantities of interest (market eigenvalues, sum of the first few sector eigenvalues and IPR of the market eigenvector) computed from the correlation matrix of a portfolio made of $N = 80$ stocks belonging to the FTSE dataset. The sampling times $T_1$ and $T_2$ are both equal to 100 days.}
\begin{ruledtabular}
\begin{tabular}{lcccccc}
Sample & $\lambda_{T_1,N}$ & $\lambda_{T_2,N}$ & $\Lambda_{T_1}$ & $\Lambda_{T_2}$ & $I_{\mathbf{v}_{T_1,N}} (\times 10^{-2})$ & $I_{\mathbf{v}_{T_2,N}} (\times 10^{-2})$ \\
\hline
    1 & 15.45 & 21.56 & 10.16 & 9.30  & 1.70  & 1.56 \\
    2 & 21.56 & 15.58 & 9.30   & 9.60   & 1.56  & 1.84 \\
    $3^*$ & 15.58 & 36.18 & 9.60   & 7.26   & 1.84  & 1.41 \\
    4 & 36.18 & 15.92 & 7.26   & 11.26 & 1.41  & 1.73 \\
    $5^*$ & 15.92 & 25.92 & 11.26 & 8.17   & 1.73  & 1.56 \\
    6 & 25.92 & 36.45 & 8.17   & 8.88   & 1.56  & 1.40 \\
    7 & 36.45 & 34.24 & 8.88   & 9.40   & 1.40  & 1.38 \\
    8 & 34.24 & 29.33 & 9.40   & 12.02 & 1.38  & 1.53 \\
    9 & 29.33 & 34.47 & 12.02 & 11.65 & 1.53  & 1.38 \\
  10 & 34.47 & 26.98 & 11.65 & 10.57 & 1.38  & 1.56 \\
  11 & 26.98 & 23.39 & 10.57 & 10.16 & 1.56  & 1.64 \\
  $12^*$ & 23.39 & 29.46 & 10.16 & 8.36   & 1.64  & 1.50 \\
  13 & 29.46 & 40.45 & 8.36   & 7.75   & 1.50   & 1.36 \\
  14 & 40.44 & 25.84 & 7.75   & 8.11   & 1.36   & 1.50 \\
  15 & 25.84 & 22.87 & 8.11   & 8.70   & 1.50   & 1.52 \\
  16 & 22.87 & 39.42 & 8.70   & 7.42   & 1.52   & 1.37 \\
\end{tabular}
\end{ruledtabular}
\end{table}
Figure \ref{portfolios} are detailed. Namely, for each of the two sampling periods $T_1$ and $T_2$ we report the values of the corresponding correlation matrix's largest eigenvalue (\emph{i.e.} the market eigenvalues $\lambda_{T_1,N}$ and $\lambda_{T_2,N}$), the sum of the next three largest eigenvalues (\emph{i.e.} the first sector eigenvalues), which we denote as $\Lambda_{T_1} = \sum_{i=N-3}^{N-1} \lambda_{T_1,i}$ and $\Lambda_{T_2} = \sum_{i=N-3}^{N-1} \lambda_{T_2,i}$, and the IPRs $I_{\mathbf{v}_{T_1,N}}$ and $I_{\mathbf{v}_{T_2,N}}$ of the eigenvectors corresponding to the largest eigenvalues. In both Tables, those samples where violations of the non-optimality bounds occur are highlighted with an asterisk $^*$. A careful inspection of the spectral quantities reported in Tables \ref{SP_portfolio} and \ref{FTSE_portfolio} allows to identify the co-occurrence of three specific phenomena in all such cases:
\begin{itemize}
	\item A very large relative increase of the market eigenvalue.
	\item A decrease of the sum of the first few sector eigenvalues.
	\item A  decrease of the IPR of the eigenvector corresponding to the market eigenvalue (market eigenvector).
\end{itemize}
In Table \ref{relvar} we report the relative variations of such quantities for the samples already highlighted in Tables \ref{SP_portfolio} and \ref{FTSE_portfolio}. Also, for both the S$\&$P and FTSE datasets, we report the largest
\begin{table}
\caption{\label{relvar} Relative changes of the spectral quantities shown in Tables \ref{SP_portfolio} and \ref{FTSE_portfolio}. Such values are reported for the highlighted samples of Tables \ref{SP_portfolio} and \ref{FTSE_portfolio}, corresponding to the ones violating the non-optimality bounds (see Figure \ref{portfolios}). Also, the largest positive variation in the market eigenvalue, and the largest negative variation in the sum of sector eigenvalues and IPR of the market eigenvalues are shown for the remaining (\emph{i.e.} non-highlighted) samples.}
\begin{ruledtabular}
\begin{tabular}{lccc}
S$\&$P Sample & $(\lambda_{T_2,N}-\lambda_{T_1,N})/\lambda_{T_1,N}$ & $(\Lambda_{T_2}-\Lambda_{T_1})/\Lambda_{T_1}$ & $(I_{\mathbf{v}_{T_2,N}}-I_{\mathbf{v}_{T_1,N}})/I_{\mathbf{v}_{T_1,N}}$ \\
\hline
$4^*$ & 54.9\% & -9.2\% & -2.1\% \\
$6^*$ & 57.7\% & -27.0\% & -14.8\% \\
Others & 42.0\% & -26.8\% & -7.5\% \\
\hline \hline
FTSE Sample & $(\lambda_{T_2,N}-\lambda_{T_1,N})/\lambda_{T_1,N}$ & $(\Lambda_{T_2}-\Lambda_{T_1})/\Lambda_{T_1}$ & $(I_{\mathbf{v}_{T_2,N}}-I_{\mathbf{v}_{T_1,N}})/I_{\mathbf{v}_{T_1,N}}$ \\
$3^*$ & 132.2\% & -24.4\% & -23.4\% \\
$5^*$ & 62.8\% & -27.4\% & -9.8\% \\
$12^*$ & 26.0\% & -17.7\% & -8.5\% \\
Others & 72.4\% & -14.7\% & -10.4\% \\
\end{tabular}
\end{ruledtabular}
\end{table}
positive variation of the market eigenvalue, and the largest negative variations of the sector eigenvalue contribution and of the market eigenvector's IPR for the non-highlighted samples (labeled as ``Others'' in Table \ref{relvar}). As can be seen, in all of the highlighted samples the aforementioned combination of relative variations is actually detected. Also, quite interestingly, the most pronounced variations, occurring for sample 3 of the FTSE dataset, appear to be responsible for the spectacular bound violation shown in the right plot of Figure \ref{portfolios}, leading to a realized risk 40 times larger than the corresponding in-sample risk.

Looking at the largest (positive or negative depending on the quantities being considered) relative variations occurring for samples other than the highlighted ones in Table \ref{relvar} shows that samples leading to severe violations of the non-optimality boundaries are not necessarily characterized by the largest overall variations. As a matter of fact, it is of paramount importance to stress that only the \emph{co-occurrence} of large relative variations for the three spectral quantities mentioned earlier (the market eigenvalue, the sum of the first few sector eigenvalues and the market eigenvector's IPR) leads to the violation of the non-optimality bounds. From a financial viewpoint, a possible interpretation of such evidence is the following one. A sudden increase of the market eigenvalue going from the sampling period $T_1$ to the sampling period $T_2$, combined with the decrease of the first few sector eigenvalues, means that approximating the PCA equation \eqref{PCA} to just one principal component over period $T_2$, \emph{i.e.}

\begin{equation} \label{PCAapprox}
r_{T_2,i}(t) \sim \sqrt{\lambda_{T_2,N}} \ v_{T_2,N}^{(i)} \ e_{T_2,N}(t),
\end{equation}
becomes a much more reasonable assumption with respect to $T_1$. Moreover, the increased delocalization of the market eigenvector over $T_2$ justifies the further approximation $v_{2,N}^{(i)} \sim  1/\sqrt{N}$, meaning, according to equation \eqref{PCAapprox}, that all stocks essentially follow the very same time evolution. This very rough, yet meaningful, approximation depicts all stocks as very strongly correlated, all equally driven by the market mode $e_{T_2,N}$. Correspondingly, all market sectors disappear. Clearly, such a picture would prevent from any chance of limiting risk, since a portfolio made by stocks evolving according to equation \eqref{PCAapprox} is absolutely unprotected against collective downwards stock movements. So, all in all, the most relevant violations of the non-optimality bounds shown in Figure \ref{portfolios} happen whenever non-stationarities in the stock dynamics cause the portfolio correlation matrix's largest eigenvalue to increase considerably and its eigenvector to delocalize almost completely, spreading evenly across all stocks. 

However, it is mandatory to notice that there does not seem to exist a cause-effect relationship between the changes in spectral quantities we mentioned and the violations of the non-optimality bounds for portfolio risk. As a matter of fact, just by carefully looking at Tables \ref{SP_portfolio} and \ref{FTSE_portfolio} one can identify a few cases (the most notable ones being sample 9 of the S$\&$P dataset and sample 16 of the FTSE dataset) where all of the previously discussed changes in spectral quantities actually occur, and the $q$ ratio between realized and in-sample risk remains well within the expected region. So, loosely speaking, large relative changes in the market eigenvalue, in its eigenvector's localization properties, and in the first few sector eigenvalues appear to be necessary but not sufficient conditions for a large spike of the $q$ ratio to take place. 

%%%%%%%%%%%%%%%%%%%%%%%%%%%%%%%%%%%%%%%%%%%%%%%%%%%%%
%%%%%%%%%%%%%%%%%%%%%%%%%%%%%%%%%%%%%%%%%%%%%%%%%%%%%
\section{Conclusions}
\label{concl}
%%%%%%%%%%%%%%%%%%%%%%%%%%%%%%%%%%%%%%%%%%%%%%%%%%%%%
%%%%%%%%%%%%%%%%%%%%%%%%%%%%%%%%%%%%%%%%%%%%%%%%%%%%%

The main goal of this paper was to study the effect of non-stationarities in financial time series on correlation coefficient measurements. The first part of this work was devoted to checking whether the common sense assumption that using longer time series provides more accurate Pearson correlation estimates actually matches financial empirical evidence. We verified that such common knowledge can be highly misleading from two different viewpoints. First, we performed a global test by checking whether correlation estimates between the same pair of stocks over non-overlapping time windows are compatible with each other. Then, we performed a local test by checking whether the addition of new information, \emph{i.e.} extending time series with the addition of new prices, actually improves the already available correlation coefficient estimates. In both tests we found clear evidence that non-negligible fractions of stock pairs and correlation estimates do not behave as intuition would predict.

The second part of this work was then devoted to study the possible effects that relying on correlation estimates computed from non-stationary price processes might have on optimal portfolio selection. We relied on the very solid framework established in \cite{PK,PKN}, where the effects of real-life portfolio selection were studied, highlighting the possible portfolio risk underestimations due to retaining optimal weights computed with past prices (\emph{i.e.} with past correlation estimates) over the future. Such a framework allowed us to check for further risk underestimations due to biased correlation estimates affected by non-stationarities. Also in this case we detected (in real financial data) serious violations of the non-optimality bounds provided in \cite{PK,PKN}, and we provided a possible explanation for such evidence in terms of correlation matrix spectral properties. Namely, the largest violations of the non-optimality bounds were found in correspondence with large relative variations of the market eigenvalues, the most relevant sector eigenvalues and the market eigenvector's IPR. However, as pointed out at the end of Section \ref{ops}, the co-occurrence of such conditions does not always seem to be enough to cause violations of the non-optimality bounds of portfolio risk. Thus, a further investigation of other possible causes for portfolio risk underestimation, maybe not even related to non-stationarities, is absolutely in order in the next future. 

We are aware that this paper sounds a bit negative, being limited to show what might go wrong when empirically estimating correlation coefficients between stocks without providing any ``healing recipe'' to the problems we discuss. In this respect, it is worth remarking that in this paper we limited ourselves to measuring risk in terms of portfolio variance, in order to keep things as simple as possible. 
However, it is well-established that portfolio variance actually does not represent the most informative risk measure one can think of. It will then be a very interesting continuation of this work to repeat the portfolio analyses performed in Section \ref{ops} on other, more refined, risk measures such as the Value at Risk \cite{Jorion} or the Expected Shortfall \cite{Acerbi} in order to test their robustness properties with respect to the changes in the portfolio correlation matrix spectral structure. Other future lines of research, which will be the topic of a forthcoming publication, will instead explore the possibility of understanding, and possibly predicting, financial crises in terms of correlation instabilities.

\acknowledgements

J.I. and G.L. acknowledge the Basque Center for Applied Mathematics in Derio, where part of this work was completed, for the hospitality. E.S. acknowledges financial support from the Italian grant PRIN 2009 ``Finitary and non-finitary probabilistic methods in economics'' 2009H8WPX$5\textunderscore002$.

\end{document}